\documentclass[onecolumn,nobibnotes,nofootinbib]{revtex4}

\usepackage{epsfig}
\usepackage{amsmath, amssymb}

\newcommand{\vk}{\mathbf{k}}
\newcommand{\vq}{\mathbf{q}}
\newcommand{\vx}{\mathbf{x}}
\newcommand{\vy}{\mathbf{y}}
\newcommand{\vz}{\mathbf{z}}
\newcommand{\vu}{\mathbf{u}}
\newcommand{\vv}{\mathbf{v}}
\newcommand{\vvr}{\mathbf{r}}
\newcommand{\vb}{\mathbf{b}}

\newcommand{\Tt}{\tilde{T}}
\newcommand{\abar}{\bar\alpha}
\newcommand{\rhosb}{\bar\rho_s}

\newcommand{\avg}[1]{\left\langle#1\right\rangle}

\begin{document}

\title{\Large QCD at high energy: saturation and fluctuation effects}
\author{Gr\'egory Soyez\footnote{On leave from the fundamental theoretical physics group of the University of Li\`ege.}}
\affiliation{SPhT, CEA/Saclay, Orme des Merisiers, F-91191 Gif-sur-Yvette cedex, France}
\email{g.soyez@ulg.ac.be}

\begin{abstract}
In these proceedings, I shall review the basic concepts of perturbative QCD in its high-energy limit. I shall concentrate on the approach to the unitarity limit, usually referred to as {\em saturation}, as well as on the gluon-number fluctuations the importance of which has recently been discovered. I shall explain the basic framework showing the need for those phenomena, first, from a simple picture of the high-energy behaviour, then, giving a short derivation of the equation driving this evolution. In the second part, I shall exhibit an analogy with statistical physics and show how this allows to derive {\em geometric scaling} in QCD with saturation. I shall finally consider the effects of gluon-number fluctuations on this picture and draw the physical consequences, {\em i.e.} a new scaling law, arising from those results.
\end{abstract}

\maketitle

\tableofcontents

\newpage

\section{Introduction}

The quest for the high-energy behaviour of perturbative QCD started thirty years ago, soon after QCD was proposed as the fundamental theory of strong interactions. To grasp this problem, it has been realised that saturation effects were to be considered in order to satisfy unitarity constraints. Throughout these proceedings, I shall summarise the present status of our understanding of this limit, which basically means finding an equation giving the evolution of amplitudes towards high energy and finding the properties of its solutions.

\begin{figure}[ht]
\centerline{\includegraphics[width=8.4cm]{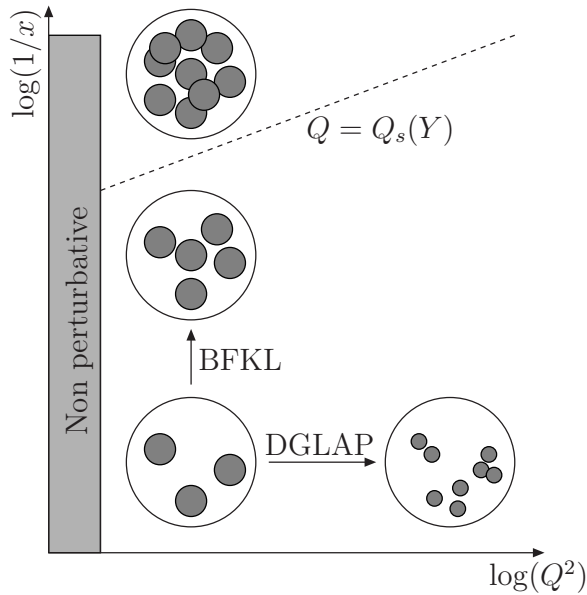}}
\caption{Picture of the proton in DIS}\label{fig:dis}
\end{figure}

In order to motivate the physical picture we want to reproduce, let us consider the problem of Deep Inelastic Scattering (DIS) {\em i.e.} $\gamma^*p\to X$ collisions. This process depends on two kinematical variables: the virtuality $Q^2$ of the photon, imposing $1/Q$ as the (transverse) resolution scale at which the photon scans the proton, and the Bjorken $x$, given by the fraction of the proton momentum carried by the parton struck by the virtual photon in a frame where the proton is moving fast. This last variable is related to the centre-of-mass energy $s=Q^2/x$, meaning that the high-energy limit corresponds to the small-$x$ limit (at fixed $Q^2$). For practical purposes, we also introduce the {\em rapidity} defined through $Y=\log(1/x)$. 

In Figure \ref{fig:dis}, we have represented the typical configuration of the proton in different domains of this phase space. We start with a proton at low $Q^2$ and low energy, represented as three partons (bottom-left part of the picture). If we increase the virtuality of the photon, we are able to resolve its partons into smaller ones. This type of evolution is described in perturbative QCD by a renormalisation group equation known as the Dokshitzer-Gribov-Lipatov-Altarelli-Parisi (DGLAP) equation \cite{dglap}. In that limit, the number of partons rises logarithmically while their typical size decreases like $1/Q^2$ so that the proton becomes more and more dilute. In what follows, we shall be interested in the other type of evolution, namely the evolution towards smaller values of $x$ for constant $Q^2$. If one boosts the proton, what typically happens is that we create new partons of a size comparable to the parents' ones. Obviously, during this evolution, known as the Balitsky-Fadin-Kuraev-Lipatov (BFKL) evolution \cite{bfkl}, the proton approaches the black-disc limit for which partons start to overlap. At that level, they start to interact among themselves and unitarity corrections are to be considered \cite{glr}. This blackening of the proton is called saturation and we shall describe it in details from perturbative QCD in these proceedings.

In addition, it has recently been discovered that gluon-number fluctuations play a crucial role in the evolution towards high energy. As we shall see later on, those fluctuations, dominating the evolution in the dilute tail ({\em i.e.} at large momentum) of the system, are amplified by BFKL evolution and thus influence greatly the whole evolution.

In the first part of these proceedings, we shall derive the equations describing the evolution towards high-energy. We shall begin our analysis with a short introduction to the dipole picture which provides a very intuitive framework to deal with high-energy QCD at large-$N_c$. This will allow for a derivation of the BFKL equation. The solution of the BFKL equation naturally leads to the introduction of unitarity corrections and to new evolution equations - the Balitsky/JIMWLK hierarchy \cite{balitsky,jimwlk} and the Balitsky-Kovchegov (BK) equation \cite{balitsky,kovchegov}. We shall then derive and discuss the contribution due to gluon-number fluctuations \cite{ms,imm,it,msw}.

In the second part, we shall consider the solutions to those evolution equations. First, we shall show that, within relevant approximations, the BK equation reduces \cite{mp} to the Fisher-Kolmogorov-Petrovsky-Piscounov (F-KPP) equation \cite{fkpp}, well studied in statistical physics. This allows for a description \cite{mp} of the asymptotic behaviour of the scattering amplitudes in terms of travelling waves, translating in QCD into the {\em geometric scaling } property \cite{geomscaling}. We shall also show that geometric scaling extends to nonzero momentum transfer \cite{bdep,bdep2}. In the last section, we shall come back to the effect of gluon-number fluctuations, which have recently proven to have important consequences \cite{gs,egm,strong,bd,bdmm,himst,msx,ims} on the approach to saturation.

\section{Evolution towards high energy}

\subsection{Small-$x$ gluons and the dipole picture}

In order to clearly emphasise the need for a resummation at high energy, let us start with a fast-moving quark of momentum $p$. This quark emits Bremsstrahlung gluons characterised by their transverse momentum $\vk_\perp^2$ and their longitudinal momentum $k_z=x p$. The probability for emitting such a gluon is
\[
dP \propto \alpha_s \frac{dx}{x}\,\frac{d\vk_\perp^2}{\vk_\perp^2}.
\]
When we consider large transverse momentum (high $Q^2$ and fixed $x$ in DIS), the collinear divergence $d\vk_\perp^2/\vk_\perp^2$ needs to be resummed and this gives rise to a renormalisation group equation known as the DGLAP equation. Throughout this proceedings, we shall instead consider the limit of high energy, in which, we are sensitive to the gluons of small longitudinal momentum $x\ll 1$ at fixed $Q^2$. This leaves a large phase-space to allow for successive soft gluon emissions satisfying $x \ll x_1 \ll \dots \ll x_n \ll 1$ which gives a contribution of order
\[
\alpha_s^n \int_x^1\frac{dx_1}{x_1}\,\dots\,\int_{x_{n-1}}^1\frac{dx_n}{x_n} = \frac{1}{n!}\alpha^n \log(1/x)^n.
\]
Even when the coupling $\alpha_s$ is small enough to ensure applicability of perturbation theory, for sufficiently small values of $x$ ({\em i.e.} at sufficiently high energy), we have $\alpha_s\log(1/x)\sim 1$ and all these contributions have to be resummed. 

\begin{figure}[ht]
\centerline{\includegraphics[width=8.4cm]{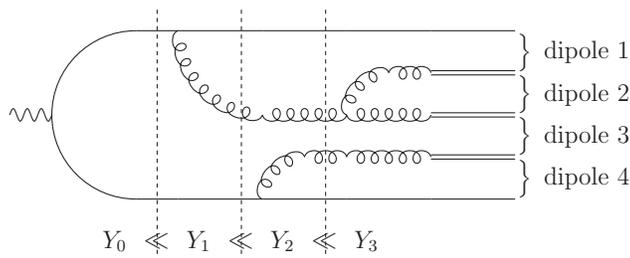}}
\caption{Onium wavefunction}\label{fig:dip_wf}
\end{figure}

In practice, it is more convenient to start with a quark-antiquark pair of transverse coordinates $\vx$ and $\vy$. Successive soft gluon emissions can be considered as independent and are depicted in figure \ref{fig:dip_wf}. In order to simplify the discussion, we shall consider the large-$N_c$ limit ($N_c$ is the number of colours). In that limit, a gluon of transverse coordinate $\vz$ can be considered as a quark-antiquark pair at point $\vz$. This means that, instead of considering a wavefunction made of quarks, antiquarks and gluons, it is sufficient to consider an {\em onium} which is an evolving system of colourless $q\bar q$ dipoles. We can thus think only in terms of dipoles. This is the dipole picture introduced by Mueller \cite{mueller}.

\begin{figure}[ht]
\centerline{\includegraphics[width=8.4cm]{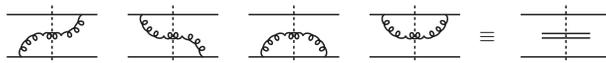}}
\caption{Gluon emission in the dipole picture is equivalent to dipole splitting}\label{fig:dip_split}
\end{figure}

\begin{figure}[ht]
\centerline{\includegraphics[width=10.4cm]{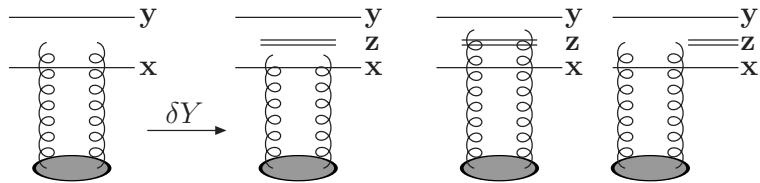}}
\caption{BFKL evolution through dipole splitting in the projectile}\label{fig:bfkl}
\end{figure}

At high-energy, all possible gluon emissions from one dipole are thus tantamount to this dipole splitting into two child dipoles as depicted in figure \ref{fig:dip_split}. We see that the probability density for a dipole of transverse coordinates $(\vx,\vy)$ to split into two child dipole $(\vx,\vz)$ and $(\vz,\vy)$ can be computed from gluon emissions from the quark and antiquark lines. The computation is performed in old-fashioned light-cone perturbation theory but we shall simply give the final result\footnote{The first equality shows explicitly the two contributions coming from emission by the quark and the antiquark.}:
\begin{equation}
dP = \frac{\abar}{2\pi}\,{\cal{M}}_{\vx\vy\vz}\, dY\,d^2z
\end{equation}
with $\abar=\alpha_s N_c/\pi$ and 
\begin{equation}
{\cal M}_{\vx\vy\vz} 
  = \left[\frac{\vx-\vz}{(\vx-\vz)^2}-\frac{\vy-\vz}{(\vy-\vz)^2}\right]^2
  = \frac{(\vx-\vy)^2}{(\vx-\vz)^2(\vz-\vy)^2}.
\end{equation}

\subsection{The BFKL equation}

Let us denotes by $\avg{T_{\vx\vy}}$ the scattering amplitude\footnote{The notation $\avg{\cdot}$ denotes the average over all possible realisations of the target colour field. It can be understood as the expectation value for the $T$-matrix operator over the target wavefunction.} for a dipole made of a quark at transverse coordinate $\vx$ and an antiquark at $\vy$. Getting the evolution of $\avg{T}$ from the dipole picture is pretty straightforward. Indeed, if one boosts the dipole $(\vx,\vy)$ from a rapidity $Y$ to a rapidity $Y+\delta Y$, this dipole shall split into two dipoles $(\vx,\vz)$ and $(\vz,\vy)$. Each of those dipoles can interact with the target which leads to the following evolution equation\footnote{The last term, coming with a minus sign, corresponds to virtual corrections}
\begin{equation}\label{eq:bfkl}
\partial_Y\avg{T_{\vx\vy}} = \frac{\abar}{2\pi} \int_z {\cal M}_{\vx\vy\vz}
                              \left(\avg{T_{\vx\vz}}+\avg{T_{\vz\vy}}-\avg{T_{\vx\vy}}\right)
\end{equation}
where we have explicitly used the probability density for dipole splitting. Equation \eqref{eq:bfkl} is the BFKL equation, derived in the mid-seventies \cite{bfkl} by Balitsky, Fadin, Kuraev and Lipatov. 

Note that here we have derived the equation by putting the evolution in the projectile dipole. It is also possible to write the evolution equation for the dipole density $n_{\vx\vy}$ in the target and to obtain \eqref{eq:bfkl} by letting the projectile dipole interact with the evolved target. We sketch out the derivation of the BFKL equation for the dipole wavefunction of the target in Appendix \ref{app:bfkln}.

\begin{figure}[ht]
\centerline{\includegraphics{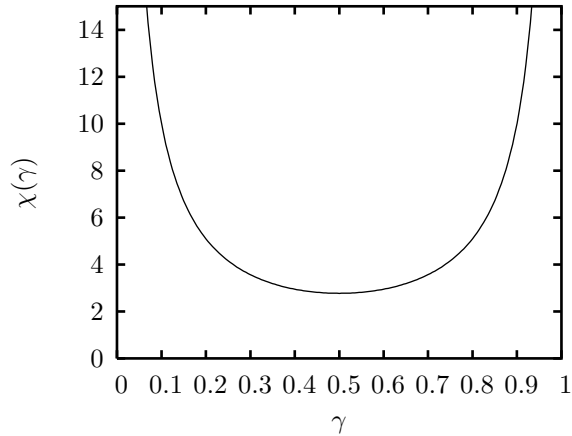}}
\caption{The BFKL eigenvalues}\label{fig:chi}
\end{figure}

This equation being linear, one expects its solution to grow exponentially. To be more explicit, if one neglects impact-parameter dependence, {\em i.e.} assumes that $\avg{T_{\vx\vy}}=\avg{T(r=|\vx-\vy|)}$, the solution of \eqref{eq:bfkl} is found by Mellin transform:
\begin{equation}\label{eq:solbfkl}
\avg{T(r)} = \int \frac{d\gamma}{2i\pi}\,T_0(\gamma)\,e^{\chi(\gamma)Y-\gamma \log(r_0^2/r^2)},
\end{equation}
where $\chi(\gamma)=2 \psi(1)-\psi(\gamma)-\psi(1-\gamma)$ is the BFKL eigenvalues plotted on Figure \ref{fig:chi}, and $T_0(\gamma)$ describes the initial condition. For a fixed dipole size, one finds the high-energy behaviour by expansion around the saddle point $\gamma={\scriptstyle{1/2}}$. One obtains
\[
\avg{T(r)} \sim \frac{r}{r_0}\exp\left[\omega_P Y-\frac{\log^2(r_0^2/r)}{2\chi''({\scriptstyle{1/2}})\abar Y} \right],
\]
with $\omega_P=4\pi\log(2)\abar$ being the BFKL Pomeron intercept. This expression suffers from two major problems: 
\begin{itemize}
\item first, even if one start with an initial condition peaked around a small value of $r$, when rapidity increases, the amplitude starts to diffuse to the large dipole sizes {\em i.e.} to the non-perturbative domain;
\item second, the solution of the BFKL equation grows exponentially with rapidity. Hence it violates the unitarity constraint $T\le 1$ obtained from first principles.
\end{itemize}

\subsection{Saturation and the BK equation}
\begin{figure}
\centerline{\includegraphics{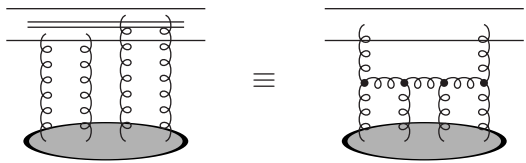}}
\caption{Multiple scattering between a target and a projectile made of two dipoles. This can be interpreted as Pomeron merging in the target.}\label{fig:multiple}
\end{figure}

This unitarity problem is precisely the point where we meet the requirement to take into account multiple interactions. The most straightforward way to see this is to come back to the splitting of a dipole $(\vx,\vy)$ into $(\vx,\vz)$ and $(\vz,\vy)$. When the target becomes dense enough, both dipoles can scatter on it. We thus have to take into account this contribution, represented in figure \ref{fig:multiple}. This leads to a quadratic suppression term in the evolution equation which becomes
\begin{equation}\label{eq:balitsky}
\partial_Y\avg{T_{\vx\vy}} = \frac{\abar}{2\pi} \int_z {\cal M}_{\vx\vy\vz}
                              \left(\avg{T_{\vx\vz}}+\avg{T_{\vz\vy}}-\avg{T_{\vx\vy}}-\avg{T^{(2)}_{\vx\vz;\vz\vy}}\right).
\end{equation}
The first thing to remark is that this new term is of the same order as the previous ones when $T^2\sim T$ or, equivalently, when $T\sim 1$. It is thus, as expected, a mandatory contribution near the unitarity limit. However, equation \eqref{eq:balitsky} involves a new object, namely $\avg{T^2_{\vx\vz;\vz\vy}}$ which probes correlations inside of the target. In a general framework, one should then write down an equation for $\avg{T^2}$, which will involve $\avg{T^2}$ through BFKL-like contributions and $\avg{T^3}$ from unitarity requirements. This ends up with a complete hierarchy, giving the evolution for each $\avg{T^k}$, known as the (large-$N_c$) Balitsky hierarchy \cite{balitsky}.

If the target is sufficiently large and homogeneous, one can simply assume $\avg{T^2_{\vx\vz;\vz\vy}}= \avg{T_{\vx\vz}}\avg{T_{\vz\vy}}$, ending up with a closed equation for $\avg{T}$
\begin{equation}\label{eq:bk}
\partial_Y\avg{T_{\vx\vy}} = \frac{\abar}{2\pi} \int_z {\cal M}_{\vx\vy\vz}
                             \left(\avg{T_{\vx\vz}}+\avg{T_{\vz\vy}}-\avg{T_{\vx\vy}}-\avg{T_{\vx\vz}}\avg{T_{\vz\vy}}
                             \right).
\end{equation}

This last expression is the Balitsky-Kovchegov (BK) equation \cite{balitsky, kovchegov}, which is the most simple equation one can obtain from perturbative QCD by including both the BFKL contributions at high-energy and the corrections from unitarity. Remembering that the dipole splitting is equivalent to gluon emission, we see from the diagram in figure \ref{fig:multiple} that the BK equation resums {\em fan diagrams} in addition to BFKL ladders. It is easy to check that $\avg{T}=0$ is an unstable fix point of the BK equation, while $\avg{T}=1$ is a stable fix point, ensuring unitarity is satisfied. 

\begin{figure}
\centerline{\includegraphics[scale=0.7]{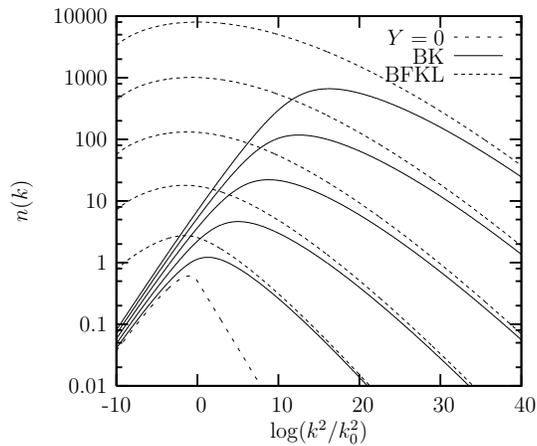}}
\caption{Dipole occupation number shown as a function of the dipole momentum $k$ for $Y=0,4,8,12,16,20$. The dashed lines correspond to the solution of the BFKL equation while the solid lines correspond to BK evolution.}\label{fig:dipdens}
\end{figure}

In addition, this new term also solves the problem of infrared diffusion. This can be seen on Figure \ref{fig:dipdens} which shows the dipole occupation number for numerical solutions of the BFKL and BK equations in momentum space ($k$ is the dipole momentum, canonically conjugated to its size). The BFKL solution exhibits the Gaussian shape predicted from \eqref{eq:solbfkl}, extending both to the infrared and to the ultraviolet. In the BK solution, we clearly see that the infrared evolution is cut and the maximum of the distribution provides a natural scale $Q_s(Y)$, called the {\em saturation momentum}, increasing with rapidity. In general, one can say that saturation corrections cut the emission of dipoles of sizes larger that the inverse of the saturation momentum. We shall come back with more details to the study of the solution of the BK equation in the next section.

\subsection{Fluctuations}

\begin{figure}
\includegraphics{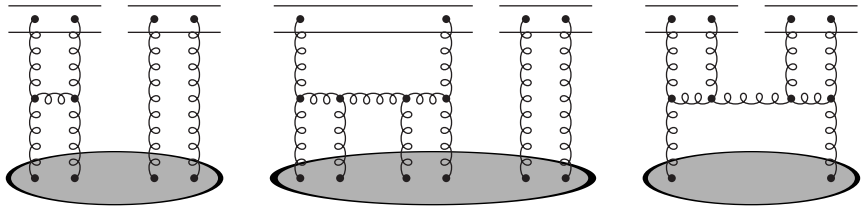}
\caption{Diagrams contributing to the evolution of $\avg{T^2}$. They correspond, from left to right, to BFKL ladders, saturation corrections and fluctuations effects.}\label{fig:t2}
\end{figure}

Very recently, it has been realised that the question of the high-energy limit of QCD, even at the leading logarithmic accuracy, is not fully described by the Balitsky hierarchy. To see this more precisely, let us consider in more details the evolution equation for $\avg{T^2}$. It corresponds to the scattering of two dipoles off a target and the diagrams contributing to one step of rapidity evolution are shown in figure \ref{fig:t2}. The first graph on the left of this figure is simply the BFKL-ladder contribution for which $\partial_Y\avg{T^2}\propto \avg{T^2}$. The diagram in the middle corresponds to multiple scattering (see also figure \ref{fig:multiple}). It accounts for unitarity corrections and gives a contribution $\partial_Y\avg{T^2}\propto \avg{T^3}$. If those two graphs were the only relevant ones in the evolution, one would obtain the (large-$N_c$) Balitsky equation. Nonetheless, it has been argued that a third contribution, represented by the rightmost diagram of figure \ref{fig:t2}, has to be included. This graph describes multiple scattering off the projectile or, equivalently, gluon-number fluctuations inside the target. After the Pomeron-merging term, contributing to saturation, one thus now also includes the Pomeron-splitting contribution, which gives rise to a new term in the evolution equation $\partial_Y\avg{T^2}\propto \avg{T}$. By combining Pomeron mergings and Pomeron splittings it is possible to build {\em Pomeron loops}. For that reason, in what follows we shall refer to those equations as the {\em Pomeron-loop equations}.

\begin{figure}
\centerline{\includegraphics{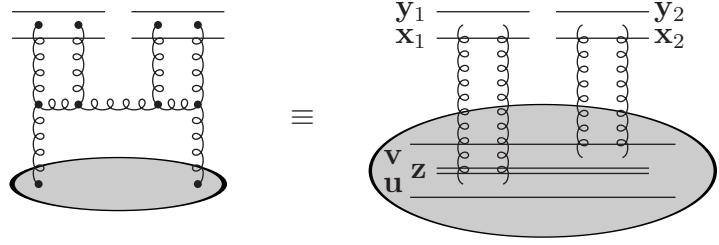}}
\caption{For a dilute target, the fluctuation diagram can be obtained from dipole splitting in the target.}\label{fig:fluct}
\end{figure}

For large-$N_c$ and at the two-gluon-exchange level, this new term has been computed \cite{it}. We shall only give the main steps hereafter. We start with the observation that gluon-number fluctuations are expected to be important when the target is still dilute, which is always the case for sufficiently small dipole sizes. In that case, it can be understood as a superposition of dipoles and the $s$-channel gluon in the fluctuation diagram (last diagram of figure \ref{fig:t2}) can be considered as emitted through a dipole splitting in the target, as represented on figure \ref{fig:fluct}.
Since we work in the two-gluon-exchange approximation, the scattering amplitude and the dipole density are related through
\begin{equation}\label{eq:ntoT}
\avg{T_{\vx\vy}} = \alpha_s^2\int_{\vu\vv} {\cal A}_0(\vx\vy|\vu\vv)\,\avg{n_{\vu\vv}}.
\end{equation}
with ${\cal A}_0$ being the dipole-dipole amplitude at the two-gluon-exchange level
\begin{equation}\label{eq:A0}
{\cal A}_0(\vx\vy|\vu\vv) = \frac{1}{8}\log^2\left[\frac{(\vx-\vu)^2(\vy-\vv)^2}{(\vx-\vv)^2(\vy-\vu)^2} \right].
\end{equation}
This relation can be inverted to obtain
\[
\avg{n_{\vx\vy}}=\alpha_s^{-2}\,\nabla_\vx^2 \nabla_\vy^2\avg{T_{\vx\vy}}.
\]
The new contribution to the evolution of $\avg{T^2_{\vx_1\vy_1;\vx_2\vy_2}}$ is thus
\begin{eqnarray}\label{eq:fluct}
\lefteqn{\left.\partial_Y\avg{T^2_{\vx_1\vy_1;\vx_2\vy_2}}\right|_{\text{fluct}}
 = \frac{1}{2}\frac{\abar}{2\pi}\left(\frac{\alpha_s}{2\pi}\right)^2}\\
&& \int_{\vu\vv\vz} {\cal M}_{\vu\vv\vz} {\cal A}_0(\vx_1\vy_1|\vu\vz){\cal A}_0(\vx_2\vy_2|\vz\vv) \nonumber
   \nabla_\vu^2 \nabla_\vv^2 \avg{T_{\vu\vv}} + (1 \leftrightarrow 2),
\end{eqnarray}

The correction \eqref{eq:fluct} becomes of the same order as the BFKL contribution when $T^2\sim \alpha_s^2 T$, or $T\sim \alpha_s^2$, {\em i.e.} in the dilute regime where, as expected, fluctuations should lead to important effects. 
Although this may at first sight seems irrelevant for the physics of saturation, one has to realise that, due to colour transparency, the amplitude becomes arbitrarily small for small dipole size, hence, even for dense targets, there is always some region of the phase-space where the system is dilute and its evolution governed by fluctuation effects.

Let us give an additional argument in favour of the importance of the fluctuation term on high-energy evolution. Once we have a Pomeron splitting coming from a fluctuation, it grows, from BFKL evolution, like two Pomerons, {\em i.e.} $T\sim\alpha_s^2\exp(2\omega_P Y)$. This has to be compared with the one-Pomeron exchange $T\sim \exp(\omega_P Y)$. We see that at very large energies $Y\ge \omega_P^{-1}\log(1/\alpha_s^2)$, BFKL evolution compensates the extra factor of $\alpha_s^2$ coming from the initial fluctuation. We shall see more precisely in the last section that those fluctuation effects in the dilute tail have important consequence on the saturation physics.

So far in this section, we have only discussed the evolution equation for $\avg{T^2}$. What we really get is an infinite hierarchy, where the evolution of $\avg{T}$ is given by \eqref{eq:balitsky}, as in the Balitsky hierarchy, and the evolution of $\avg{T^k}$ contains three contributions: a linear BFKL term proportional to $\avg{T^k}$, a (negative) saturation term  proportional to $\avg{T^{k+1}}$ as it appears in the Balitsky hierarchy, and the new fluctuation contribution proportional to $\avg{T^{k-1}}$, similar to \eqref{eq:fluct}. Remarkably enough, it is possible to show that this infinite hierarchy for the average amplitudes can be rewritten as a single Langevin equation\footnote{We show the equivalence between the two formulations in appendix \ref{app:langevin}, for a simpler of Langevin equation encountered in Section 4.}
\begin{eqnarray}\label{eq:langevin_full}
\partial_Y T_{\vx\vy} & = & \frac{\abar}{2\pi} \int_\vz\,{\cal M}_{\vx\vy\vz} 
  \left[ T_{\vx\vz} + T_{\vz\vy} - T_{\vx\vy} - T_{\vx\vz}T_{\vz\vy}\right]\\\nonumber
& + & \frac{1}{2} \frac{\abar}{2\pi}\frac{\alpha_s}{2\pi} \int_{\vu\vv\vz} {\cal{A}}_0(\vx\vy|\vu\vz) \frac{|\vu-\vv|}{(\vu-\vz)^2} \sqrt{\nabla^2_\vu\nabla^2_\vv T_{\vu\vv}}\, \nu_{\vu\vv\vz;Y},
\end{eqnarray}
where $\nu$ is a Gaussian white noise satisfying
\begin{eqnarray*}
\avg{\nu_{\vu\vv\vz;Y}} & = & 0,\\
\avg{\nu_{\vu\vv\vz;Y}\,\nu_{\vu'\vv'\vz';Y'}} 
 & = & \delta(\abar Y-\abar Y') \delta^{(2)}(\vu-\vv')  \delta^{(2)}(\vz-\vz')\delta^{(2)}(\vv-\vu').
\end{eqnarray*}
Different realisations of the noise term lead to different events $T_{\vx,\vy}$, from which one can compute $\avg{T^k}$ using standard statistical techniques. However, dealing with equation \eqref{eq:langevin_full} remains very difficult since the noise term is off-diagonal and non-local.

Finally, let us conclude this section by quoting that the equation \eqref{eq:fluct}, including linear, saturation and fluctuation effects is the most complete equation known in perturbative high-energy QCD so far. Its extension beyond the large-$N_c$ approximation is still a challenging problem.

\subsection{QCD as a reaction-diffusion process ?}

Let is make a short digression and consider a {\em reaction-diffusion} system in which we have a set of particles evolving through splitting at a rate $\gamma$ and merging at a rate $\sigma$:
\[
A\stackrel{\gamma}{\underset{\sigma}{\rightleftharpoons}}A+A.
\]
For simplicity, we shall not consider any spatial dimension, though it only introduces technical difficulties without altering the conclusions. If we want to study the time evolution of such a system, the best way is to start with the {\em master equation} which gives the evolution of $P_n$, the probability to have $n$ particles. It contains four terms, corresponding to gain and loss terms for splitting and merging weighted by adequate combinatorial factors:
\[
\partial_t P_N = \underbrace{\gamma\, (N-1)P_{N-1}}_{\text{gain}}
               - \underbrace{\gamma\, NP_N}_{\text{loss}}
               + \underbrace{\sigma\, N(N+1)P_{N+1}}_{\text{gain}}
               - \underbrace{\sigma\, N(N-1)P_N}_{\text{loss}}
\]

In practice, this equation is only helpful to find evolution of observable, averaged, quantities. The most simple example is the average particle number
\[
\avg{n} = \sum_{N=0}^\infty N\,P_N,
\]
which can be generalised to a subset of $k$ particles
\[
\avg{n^k} \equiv \sum_{N=k}^\infty \frac{N!}{(N-k)!}\,P_N.
\]
Using the master equation, one finds the evolution of the particle number and its correlators after a bit of algebra
\[
\partial_t \avg{n^k} = \gamma\, k\avg{n^k} + \gamma\, k(k-1)\avg{n^{k-1}}
                     - \sigma\, k(k+1) \avg{n^{k+1}} - \sigma\, k\avg{n^k}.
\]

More interesting is the evolution of the scattering amplitude for this system off a generic target, defined by
\[
{\cal A}(t) = \sum_{k=0}^\infty (-)^k\:\avg{n^k}_{t_0}\:\avg{T^k}_{t-t_0},
\]
where we evolve the particle system up to a time $t_0$ and the scattering matrix for the target from $t_0$ to $t$. To be consistent, this definition should not depend on any specific choice for $t_0$. This gives a renormalisation group equation from which one can infer the evolution of the scattering amplitude:
\[
\partial_t \avg{T^k} = \underbrace{\gamma \avg{T^k}}_{\text{BFKL}}
                     - \underbrace{\gamma \avg{T^{k+1}}}_{\text{sat.}}
                     + \underbrace{\sigma \left[\avg{T^{k-1}}-\avg{T^k}\right]}_{\text{fluct.}}.
\]
We clearly recognise in this expression the BFKL, saturation and fluctuations contributions obtained in QCD\footnote{The second fluctuation term disappears in the case of QCD.}. 

Inserting the correct spatial degrees of freedom to account for transverse coordinates, we find \cite{ist} that the high-energy QCD evolution hierarchy obtained in the previous section can be seen as an {\em effective dipole reaction-diffusion process} where rapidity plays the role of time and with the vertices
\begin{eqnarray}
\gamma(\vx\vy\to\vx\vz,\vz\vy) & \sim & \abar{\cal{M}}_{\vx\vy\vz},\\
\sigma(\vx_1\vy_1,\vx_2\vy_2\to\vu\vv)
 & \sim & \abar\alpha_s^2 \nabla_\vu^2\nabla_\vv^2\left\lbrack
          {\cal M}_{\vu\vv\vz}{\cal A}_0(\vx_1\vy_1|\vu\vz){\cal A}_0(\vx_2\vy_2|\vz\vv)
          \right\rbrack
\end{eqnarray}
for splitting and merging respectively. One recognise in these expressions that the splitting is given by the BFKL kernel while the merging comes from the computation in the previous section. 

This interesting analogy is however to be taken with care. Indeed, the merging vertex is not positive defined. Physically, this comes from the fact that fluctuations really involve the gluonic degrees of freedom. Their interpretation as a reaction-diffusion system in terms of dipoles can hence only be effective.

\section{Properties of the BK scattering amplitudes}

In this section, we shall sketch out the main properties of the solutions of the evolution equations towards high energy. As we shall explain in details in the next lines the major source of information comes from the analogy between the BK equation and the Fisher-Kolmogorov-Petrovsky-Piscounov (F-KPP) equation which is well studied in statistical physics. This section shall be devoted to derive this analogy and its consequences in the case of the BK equation. We shall first consider the impact-parameter-independent BK equation and then show how the arguments extend to the full BK equation. We leave the discussion concerning the effect of fluctuations for the next section.

\subsection{Statistical physics and geometric scaling}

So, let us start with the BK equation \eqref{eq:bk}. We shall first restrict ourselves to the impact-parameter-independent version of the equation, for which one can easily go to momentum space by using\footnote{In this section, we omit the $\avg{\cdot}$ quotations as they do not play any role in the mean-field approximation.}
\[
\Tt(k) = \frac{1}{2\pi}\int \frac{d^2r}{r^2}\,e^{i\vk.\vvr}\,T(r) 
       = \int_0^\infty \frac{dr}{r}\, J_0(kr)\, T(r),
\]
and where the equation becomes
\[
\partial_Y \Tt(k) = \frac{\bar\alpha}{\pi}\int \frac{dp^2}{p^2}
                    \left[\frac{p^2\Tt(p)-k^2\Tt(k)}{|k^2-p^2|} + \frac{k^2\Tt(k)}{\sqrt{4p^4+k^4}}\right]
                  - \bar\alpha \Tt^2(k).
\]
It is actually possible to rewrite the integral in is equation in term of a differential operator:
\begin{equation}\label{eq:bkk}
\partial_Y \Tt(k) = \abar \chi(-\partial_L)\Tt(k) - \abar \Tt^2(k),
\end{equation}
with $L = \log(k^2/k_0^2)$, $k_0$ being a soft reference scale and $\chi(\gamma)$ is the BFKL kernel. 

In order to simplify the problem, we shall work in the saddle point approximation (often referred to as the {\em diffusive approximation}) {\em i.e.} expand the BFKL kernel to second order around $\gamma=1/2$, transforming the complicated differential operator in equation \eqref{eq:bkk} into a second-order operator
\[
\chi(-\partial_L)\Tt(k) \approx \chi({\scriptstyle \frac{1}{2}}) 
 + {\scriptstyle \frac{1}{2}}\chi''({\scriptstyle \frac{1}{2}})\left(\partial_L+{\scriptstyle \frac{1}{2}}\right)^2.
\]
Up to a linear change of variable switching from $Y$ and $L$ to time $t=\abar Y$ and space $x = L + \text{cst}.Y$ and a renormalisation of the amplitude $T\to u=\text{cst}.T$, the BK equation in the saddle point approximation becomes
\[
\partial_t u(x,t) = \partial_x^2u(x,t) + u(x,t) - u^2(x,t).
\]
This equation is nothing but the F-KPP equation studied in statistical physics since forty years and applying to many problem. It describes reaction diffusion processes in the mean-field approximation where one can have creation and annihilation of particles locally and diffusion to neighbouring site. This equation has many applications {\em e.g.} in chemistry and biology.

\begin{figure}
\centerline{\includegraphics[scale=0.8]{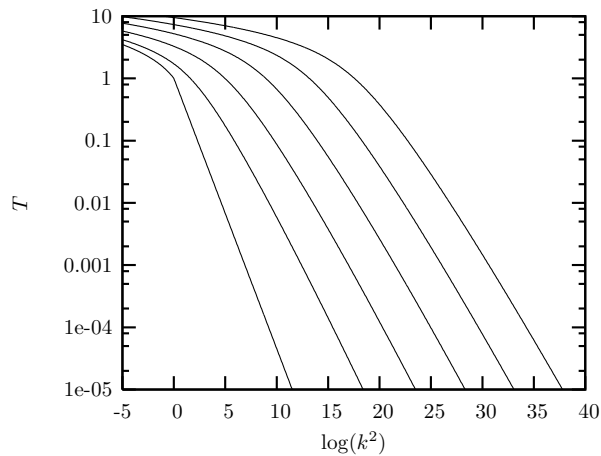}}
\caption{Numerical simulation of the rapidity-evolution of the BK equation. We start at $Y=0$ from the leftmost amplitude and evolve to higher $Y$ using \eqref{eq:bkk}. Amplitude is shown for $Y=5,10,15,20,25$ and clearly exhibits a travelling-wave pattern.}\label{fig:bkfront}
\end{figure}

One knows that the F-KPP equation admits travelling waves as asymptotic solutions {\em i.e.} there exists a critical velocity $v_c$ and a critical slope $\gamma_c$, determined only from the knowledge of the linear kernel of the F-KPP equation $\partial_x^2+\mathbf{1}$, such that (see figure \ref{fig:bkfront} for a pictorial representation coming from the BK equation which we shall explain a bit later)
\[
u(x,t) \stackrel{t\to \infty}{\longrightarrow} u(x-v_ct) \stackrel{x\gg v_ct}{\approx} e^{-\gamma_c(x-v_ct)} (x-v_ct).
\]
Although working in the diffusive approximation makes the discussion easier due to the direct matching with the F-KPP equation, this assumption is not required. Indeed, for an equation with a more complicated kernel (than $\partial_x^2+\mathbf{1}$), one can prove the existence of travelling waves provided the three following conditions are satisfied:
\begin{enumerate}
\item the amplitude has $0$ as unstable fix point and $1$ as stable fix point,
\item the initial condition decreases faster than $\exp(-\gamma_c L)$ at large $L$ (see below for the general definition of $\gamma_c$),
\item the equation obtained by neglecting the nonlinear terms admits superposition of waves as solution:
\[
\left.u(x,t)\right|_{\text{lin}} = \int \frac{d\gamma}{2i\pi}\,u_0(\gamma)\,\exp\left[-\gamma(x-v_\gamma t)\right],
\]
where the wave of slope $\gamma$ travels at a speed $v_\gamma$.
\end{enumerate}
Then, travelling waves are formed during evolution with the critical parameters $\gamma_c$ and $v_c$ obtained, from the linear kernel only, through the relation
\[
v_c=\chi'(\gamma_c)=\frac{\chi(\gamma_c)}{\gamma_c}.
\]
This corresponds to the selection of the wave with the minimal speed $\chi(\gamma)/\gamma$ and the point where group and phase velocities are equal.

\begin{figure}
\centerline{\includegraphics[scale=1.2]{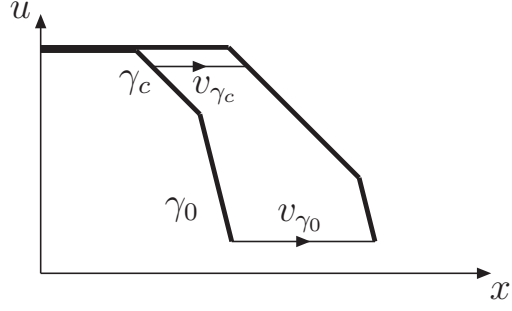}}
\caption{Formation of a travelling wave from a schematic evolution of a wavefront in time/rapidity.}\label{fig:schem}
\end{figure}

Before going any further, let us give a qualitative argument explaining how the minimal speed is selected. To simplify the discussion, we shall consider the initial front at the left of figure \ref{fig:schem}. At an initial time, it captures the main features of the spatial dependence: saturation (first condition here-above) followed by a decrease with the critical exponent $\gamma_c$ and finally a steeper decrease (second condition here-above). After one step of time evolution, each of the decreasing fronts shall evolve with its own speed (those fronts are in a dilute regime so we can use the third condition). The result of this evolution is shown on Figure \ref{fig:schem} and we clearly see that the region over which the critical slope, {\em i.e.} the one with the minimal speed, is growing during the evolution. This selection mechanism is very general. It mainly relies on the properties of the linear equation and saturation only ensures that the slope is $\gamma_c$ at some point in the initial condition.

\begin{figure}
\centerline{\includegraphics[scale=0.8]{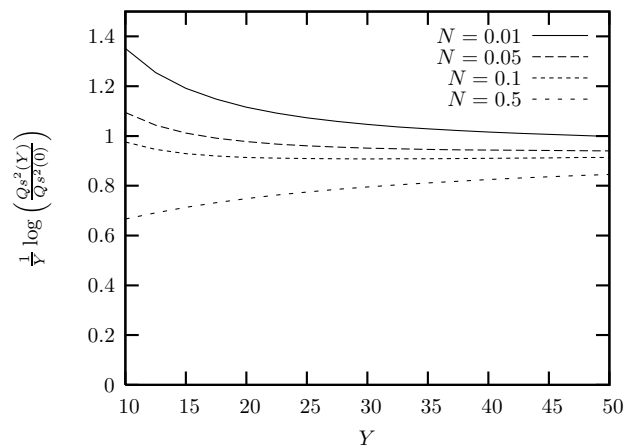}}
\caption{Rapidity evolution of the saturation scale extracted from figure \ref{fig:bkfront}. More precisely, the speed of the wave, {\em i.e.} the exponent of the saturation scale, is plotted. It goes to a constant as it should.}\label{fig:bkqs}
\end{figure}

In the case of the BK equation, the first condition is satisfied due to BFKL growth and saturation, the second comes from colour transparency ($T\propto 1/k^2=\exp(-L)$) and the last one is equivalent to the solution \eqref{eq:solbfkl} (with $r/r_0$ replaced by $k_0/k$) of the BFKL equation obtained by neglecting nonlinear terms in \eqref{eq:bkk}.

The critical parameters (for the complete BFKL kernel {\em i.e.} for \eqref{eq:bkk}), are found to be $\gamma_c\approx 0.6275$ and $v_c\approx 4.8836\abar$. Written in terms of the QCD variables $Y$ and $k^2$, one can then write the asymptotic solution for the impact-parameter-independent BK equation under the form
\begin{equation}\label{eq:bkfront}
T(k;Y)
 \stackrel{Y\to\infty}{=} T\left(\frac{k^2}{Q_s^2(Y)}\right)
 \stackrel{k\gg Q_s}{=}\left[\frac{k^2}{Q_s^2(Y)}\right]^{-\gamma_c}\log\left[\frac{k^2}{Q_s^2(Y)}\right],
\end{equation}
with the saturation scale given by
\begin{equation}\label{eq:bkqs}
Q_s^2(Y) \stackrel{Y\to\infty}{=} k_0^2\,\exp\left[ v_c Y - \frac{3}{2\gamma_c}\log(Y) \right].
\end{equation}

The formation of a travelling-wave pattern when energy increases can easily be seen on numerical simulations of the BK equation. As displayed in figure \ref{fig:bkfront}, if one start with a steep enough initial condition (leftmost curve), the amplitude increases with energy and a wave moving into the dilute domain gets formed. From that simulation, one can extract the saturation scale by solving $T(Q_s^2,Y)=N$ at each values of $Y$ and for a fixed threshold $N$. In figure \ref{fig:bkqs}, we have plotted $\log(Q_s^2)/Y$ which goes to a constant value as it is expected from \eqref{eq:bkqs}.

Equation \eqref{eq:bkfront} has a remarkable property: it proves that at high energy the amplitude, {\em a priori} a function of both $Y$ and $k$, depends only on the ratio between $k$ and the saturation momentum. This property, known as {\em geometric scaling}, has been observed \cite{geomscaling} in the HERA measurements of the proton structure function. The fact that geometric scaling can be derived from the BK equation is one of the most important indication for the experimental observation of saturation. The saturation scale obtained from the structure function data is of the order of 1 GeV for $x\sim 10^{-5}$. This means that high-energy QCD can indeed be studied from the point of view of perturbation theory.

\subsection{Nonzero momentum transfer}

Up to now, we have only discussed the impact-parameter-independent BK equation. We might therefore ask whether or not these arguments extend to the full equation including all phase-space degrees of freedom. A dipole of transverse coordinates $(\vx,\vy)$ is then represented through its size $\vvr = \vx-\vy$ and impact parameter $\vb=(\vx+\vy)/2$. The problem is then that dipole splitting is non-local in impact parameter. This means that the BK equation couples different values of $\vb$ and we can not apply directly the previous arguments for each value of $\vb$. Again, the solution consists in moving to momentum space and replace the impact parameter by the momentum transfer $\vq$
\[
\tilde T(\vk,\vq) = \int d^2x\,d^2y\,e^{i\vk.\vx}e^{i(\vq-\vk).\vy}\frac{T(\vx,\vy)}{(\vx-\vy)^2}.
\]
The BK equation then takes a form \cite{bdep} for which the BFKL kernel is local in $\vq$ (the non-locality of the non-linear term is not important for our purposes as its only role is to ensure saturation)
\begin{eqnarray}\label{eq:bkfull}
\lefteqn{\partial_Y \tilde T(\vk,\vq)}\\
& = & \frac{\bar\alpha}{\pi}\int \frac{d^2k'}{(\vk-\vk')^2}\left\{
      \tilde T(\vk',\vq) - \frac{1}{4}\left\lbrack
      \frac{\vk^2}{\vk'^2}+\frac{(\vq-\vk)^2}{(\vq-\vk')^2}
      \right\rbrack\tilde T(\vk,\vq)\right\}\nonumber\\
& - & \frac{\bar\alpha}{2\pi}\int d^2k'\,\tilde T(\vk,\vk')\tilde T(\vk-\vk',\vq-\vk').\nonumber
\end{eqnarray}

We can now proceed with this equation in a similar way as for the the impact-parameter-independent case. The existence of travelling waves requires the three conditions stated previously to be satisfied. The two first ones are trivially satisfied \footnote{Indeed, we have saturation and BFKL growth ensuring the first condition and colour transparency still apply for the second one.}. For the third condition to be valid, we need to find solutions of the linear part of \eqref{eq:bkfull}, {\em i.e.} the complete BFKL equation, which can be expressed as a superposition of waves. This consists in a careful treatment of the solutions of the BFKL equation \cite{bfklsol} including all phase-space variables. It turns out that, using the momentum transfer $\vq$ instead of the impact parameter $\vb$ , leads to a powerful factorisation between the target and the projectile (see Appendix \ref{app:nonzero} for more details). Then, we can show \cite{bdep2} that when the dipole momentum $k$ is much larger than the momentum transfer $q$ and the typical scale of the target $k_0$, the solutions of the BFKL equation are a superposition of waves and therefore, we obtain travelling waves for the full BK equation. More precisely, the high-energy behaviour of the amplitude takes the form
\begin{eqnarray*}
\tilde T(k,q;Y)
& \stackrel{Y\to\infty}{=} & T\left(\frac{k^2}{Q_s^2(q;Y)}\right)\\
& \stackrel{k\gg Q_s}{=}   & \left[\frac{k^2}{Q_s^2(q;Y)}\right]^{-\gamma_c}\log\left[\frac{k^2}{Q_s^2(q;Y)}\right].
\end{eqnarray*}
This expression is exactly the same as for the previous case except that the saturation scale now depends on momentum transfer
\[
Q_s^2(q;Y) \stackrel{Y\to\infty}{=} \Lambda^2\;\exp\left[v_c Y - \frac{3}{2\gamma_c}\log(Y)\right]
\]
with
\[
\Lambda^2 = \begin{cases}
 k_0^2 & \text{if } k_0 \gg q\\
 q^2   & \text{if } q \gg k_0
\end{cases}.
\]
We explain in Appendix \ref{app:nonzero} how the exponential behaviours required for travelling-waves formation appear and how the reference scale $k_0$ or $q$ emerges. Naively, we can expect the saturation scale at large $q$ to be determined by $q$ as, in the tail of the amplitude, we are dominated by the hard part of the scattering. In addition, when $q\to 0$ we should recover the $b$-independent result and some soft reference scale should come back.

It is very interesting to notice that the critical slope $\gamma_c$ and speed $v_c$, obtained from the BFKL kernel only, are the same as for the impact-parameter-independent case. 

The important point at this stage is that this study {\em predicts} geometric scaling at non-zero momentum transfer. Experimental measurements of the DVCS cross-sections or diffractive $\rho$-meson production are good candidates and can provide more evidence for saturation.

\section{High-energy QCD with fluctuations}

Let us now consider the effect of the fluctuation contribution \eqref{eq:fluct}. This new term, added to the full Balitsky hierarchy, turns a simple non-linear equation into an infinite hierarchy with a complicated transverse-plane dependence\footnote{The fluctuation term involves an awkward integration. By integration by part one can express it as a vertex applied to $T$ but the integration over the internal variable $\vz$ in \eqref{eq:fluct} is unknown.}. We can however simplify the equation to a tractable problem by performing a local-fluctuations approximation. This amounts to simplify the dipole-dipole scattering amplitude ${\cal A}_0$ used to relate the dipole density $n$ with the scattering amplitude $T$. We assume that two dipoles interact if they are of the same size and if their centre-of-mass are sufficiently close to allow for an overlap of the two dipoles. Within this approximation equation \eqref{eq:ntoT} is replaced by
\[
\avg{T_{\vx\vy}} = \kappa\alpha_s^2 \,\left|\vx-\vy\right|^4\,\avg{n_{\vx\vy}},
\]
where $\kappa$ is an unknown fudge factor of order 1.

Once this is done, the remaining steps are as follows: one Fourier-transform the dipole size $r=|\vx-\vy|$ into momentum $k$ and perform a coarse-graining approximation to get rid of the impact-parameter dependence of the fluctuation term. This computation results into the following simplified form of the hierarchy (we give only the expression for $\avg{T}$ and $\avg{T^2}$ for simplicity)
\begin{eqnarray*}
\partial_Y \avg{T_k} & = & \bar\alpha \chi(-\partial_L)\avg{T_k}
                         - \bar\alpha \avg{T^2_{k,k}}, \\
\partial_Y \avg{T^2_{k_1,k_2}} & = & \bar\alpha \chi(-\partial_{L_1})\avg{T^2_{k_1,k_2}}
                                   - \bar\alpha \avg{T^3_{k_1,k_1,k_2}} + (1 \leftrightarrow 2)\\
            & + & \bar\alpha \, \kappa\alpha_s^2\, k_1^2 \delta(k_1^2-k_2^2)\avg{T_{k_1}},
\end{eqnarray*}
where, as previously, $L_i = \log(k_i^2/k_0^2)$.

Again, this infinite hierarchy has the advantage that it can be rewritten under the form of a Langevin equation\footnote{As we have seen previously, the complete hierarchy using \eqref{eq:fluct} can also be rewritten as a Langevin equation \eqref{eq:langevin_full}. However, the noise term here, being local, appears to be much simpler than for the general case \eqref{eq:langevin_full}.}
\begin{equation}\label{eq:langevin}
\partial_Y T(L)
  = \bar\alpha \left[\chi(-\partial_L)T(L)-T^2(L)+\sqrt{\kappa\alpha_s^2T(L)}\eta(L,Y)\right],
\end{equation}
where $\eta$ is a Gaussian white noise satisfying the following commutation relations:
\begin{eqnarray}\label{eq:noise}
\avg{\eta(L,Y)} & = & 0,\nonumber\\[-3mm]
& & \\[-3mm]
\avg{\eta(L_1,Y_1)\eta(L_2,Y_2)} & = & \frac{4}{\bar\alpha}\delta(L_1-L_2)\delta(Y_1-Y_2). \nonumber
\end{eqnarray}

The Langevin equation is a stochastic equation describing an event-by-event picture. Each realisation of the noise term corresponds to a particular evolution of the target and one can show (see Appendix \ref{app:langevin}) that once we average over those realisations using the correlations \eqref{eq:noise} the complete hierarchy for the evolution of $\avg{T^k}$ is recovered. It is interesting to notice that equation \eqref{eq:langevin} is formally equivalent to the BK equation with an additional noise term.

As for the case of the BK equation, we can first restrict ourselves to the diffusive approximation. This leads to the stochastic FKPP (sFKPP) equation which amounts to add a noise term to the FKPP equation. It has found many applications in various fields, {\em e.g.} in the description of reaction-diffusion systems. The noise term appears once we have to consider discreteness effects ({\em e.g.} for a reaction-diffusion system on a lattice, only an integer number of particles are admitted per site). When the number of particles involved goes to infinity, the mean-field approximation seems justified and the (non-stochastic) FKPP equation is valid. 

\begin{figure}
\centerline{\includegraphics[scale=0.66]{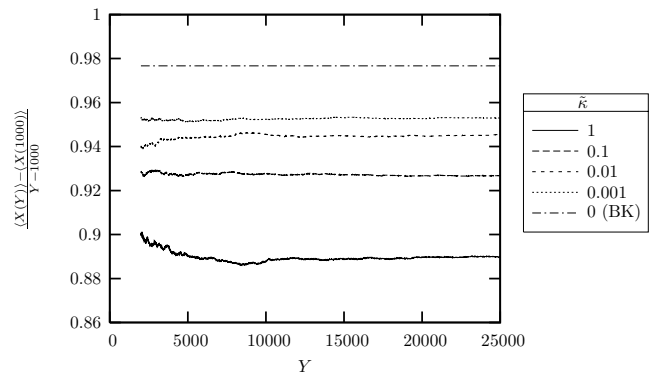}}
\caption{Speed of the wave for different values of the noise strength $\kappa\alpha_s^2$. The BK speed has been added for comparison. One sees that the speed decreases when $\kappa\alpha_s^2$ increases.}\label{fig:speed}
\end{figure}

\begin{figure}
\centerline{\includegraphics[scale=0.71]{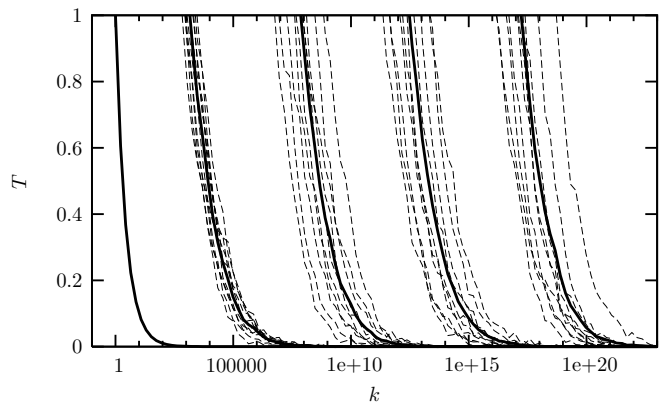}}
\caption{Numerical simulations of equation \eqref{eq:langevin}. The dashed curves are different events corresponding to the same initial condition with different realisations of the noise term. Travelling waves are observed for each curve, together with dispersion. The black curve is the average amplitude.}\label{fig:events}
\end{figure}

The basic effects of the additional noise term in the sFKPP equation are known, at least qualitatively. Despite the fact that the fluctuations are only expected to have large consequences on the dilute tail of the wavefront, these modifications change the picture at saturation. In order to test the validity of those results in the case beyond the diffusive approximation, {\em i.e.} with the full BFKL kernel, we have performed \cite{gs} numerical studies (see also \cite{egm}) of the QCD equation \eqref{eq:langevin}. Moreover, many analytical results are only known in the limit where the noise strength $\kappa\alpha_s^2$ is (irrealistically for QCD) small while we have concentrated our numerical work of physically acceptable values. In the following paragraphs, we review the main effects of the fluctuation contribution and show that they are observed in the numerical analysis.

First of all, for a single event, the evolved amplitude shows a travelling-wave pattern (up to small fluctuations in the far tail which, at least a this level, are irrelevant for the wave pattern at saturation). This means that each single realisation of the noise leads to geometric scaling, as it is the case for the BK equation. At this level, the major difference comes from a decrease of the speed of the wave. Analytically, this can only be computed in the limit $\kappa\alpha_s^2\to 0$ {\em i.e.} when the fluctuations are a small perturbation around the mean-field behaviour\footnote{A recent analysis has also computed its strong-noise behaviour \cite{strong}.}. It has been found \cite{bd,bdmm} that 
\[
v^* \underset{\alpha_s^2\kappa\to 0}{\to} v_c - \frac{\abar\pi^2\gamma_c\chi''(\gamma_c)}{2\log^2(\alpha_s^2\kappa)}.
\]
Unfortunately, this expression gives only reliable results for extremely small values of $\kappa\alpha_s^2\to 0$ such as $10^{-20}$, which is not sufficient for realistic situations. The numerical simulations we have performed clearly exhibits this decrease of the speed when the noise strength $\kappa\alpha_s^2$ increases (see figure \ref{fig:speed}).

The second noticeable effect of the noise term is to introduce dispersion between different events. Different events have the same shape but their position $X(Y)$ ($\log[Q_s^2(Y)]$ in physical variables) fluctuates. At a given rapidity, one can compute the dispersion of those events. This diffusion process being very similar to a random walk, one expects that the dispersion of the events behaves like $\sqrt{Y}$:
\[
\avg{X^2}_Y-\avg{X}_Y^2 \stackrel{Y\to\infty}{\approx} D_{\text{diff}}Y.
\]
The parametric dependence of the diffusion coefficient $D_{\text{diff}}$ has been computed numerically \cite{bd} and behaves\footnote{Recently, this coefficient has been computed analytically \cite{bdmm} for $\kappa\alpha_s^2\ll 1$.} like $|\log^{-3}(\kappa\alpha_s^2)|$ for small $\kappa\alpha_s^2$.

\begin{figure}
\centerline{\includegraphics[scale=0.67]{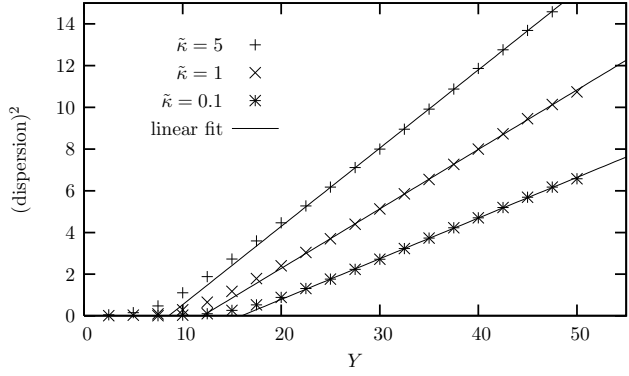}}
\caption{Dispersion (squared) of the position of the events as a function of rapidity. As expected, it dispersion increases like $\sqrt{Y}$ but few dispersion is obtained in early stages of the evolution.}\label{fig:disp}
\end{figure}

\begin{figure}
\centerline{\includegraphics[scale=0.69]{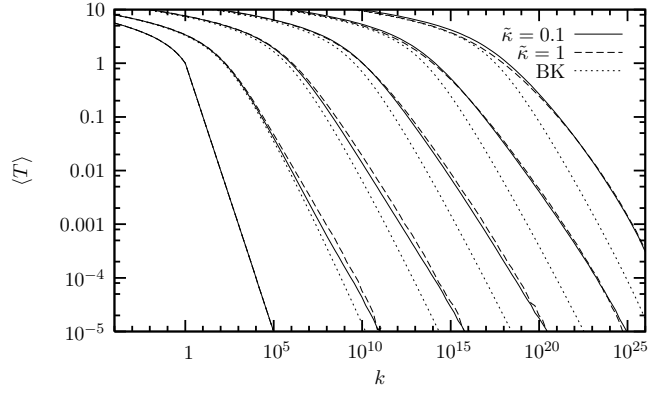}}
\caption{Evolution of the averaged amplitude for different values of the noise strength, compared with the mean field result (dotted curve). The curves correspond, from left to right, to $Y=0,10,20,30,40$ and $50$.}\label{fig:front-avg}
\end{figure}

Concerning the numerical simulations, the dispersion of the wavefront between different events and its increase with rapidity are manifest on figure \ref{fig:events}. One can then extract the dispersion as a function of rapidity. The resulting curve, shown on figure \ref{fig:disp} for different values of the noise strength, calls for two remarks. First, the expected asymptotic behaviour (dispersion $\propto \sqrt{Y}$) is observed and the diffusion coefficient $D_{\text{diff}}$ increases with rapidity and with the noise strength. However, for small values of the rapidity, we do not observe a significant dispersion. This lack of dispersion may come from the fact that we have to wait for the travelling front to get formed before dispersion becomes significant. However, those effects in the beginning of the evolution are perhaps not universal and deserve more detailed studies.

This dispersion of the events has an important physical consequence: although each single event displays geometric scaling, once we compute the average amplitude, dispersion induces geometric scaling violations. This effect is best seen on figure \ref{fig:front-avg} where we have compared the evolution including the fluctuations to the BK results. While in the BK equation, a fixed travelling-wave pattern is formed, once fluctuations are taken into account, we see a broadening of the average amplitude as rapidity increases.

Those violations of geometric scaling can be studied \cite{himst} in a model which encompasses all the physical arguments presented here-above and which shall prove to be very efficient for phenomenological studies. In order to reconstruct physical amplitudes, we need two building ingredients:
\begin{itemize}
\item a front-by-front amplitude which displays geometric scaling. This amplitude has to satisfy unitarity in the infrared and must decrease exponentially in the ultraviolet. The simplest choice, which is actually sufficient for what follows, is
\begin{equation}\label{eq:event_front}
T_{\text{event}}(\rho-\rho_s) = \begin{cases}
 \exp[-\gamma_0 (\rho- \rho_s)] & \text{if }\rho\ge\rho_s,\\
 1                              & \text{if }\rho < \rho_s,
\end{cases}
\end{equation}
where we have introduced the notations $\rho = \log(r_0^2/r^2)$ and $\rho_s = \log[1/(r^2Q_s^2)]$. This expression shows indeed geometric scaling since it is a function of $\rho-\rho_s$ only. The precise value of $\gamma_0$ is not relevant for what follows, although it might be adequate to adopt $\gamma_0=\gamma_c=0.6275$ or $\gamma_0=1$.
\item a dispersion between different events. Physically, this dispersion of the position of the events, {\i.e. of their saturation scale}, means that the saturation scale becomes a random variable. We thus need a probability distribution for $\rho_s$. We shall take a Gaussian form\footnote{We can actually prove \cite{msx} that, at high-energy, the probability distribution can be considered as Gaussian over an arbitrarily large domain.}
\begin{equation}\label{eq:proba}
P(\rho_s) = \frac{1}{\sqrt{2\pi\sigma^2}}\,\exp\left[\frac{(\rho_s-\rhosb)^2}{2\sigma^2}\right].
\end{equation}
The parameters of the Gaussian (the average saturation scale $\rhosb$ and the dispersion $\sigma^2$) are expected to grow linearly with rapidity, but we shall treat them as parameters.
\end{itemize}
The average amplitude can be constructed from \eqref{eq:event_front} and \eqref{eq:proba}
\begin{eqnarray*}
\avg{T(\rho)} & = & \int_{-\infty}^{\infty} d\rho_s\,P(\rho_s)\,T_{\text{event}}(\rho-\rho_s)\\
              & = & \frac{1}{2}{\rm erfc}\left(\frac{z}{\sqrt{2}\sigma}\right)
                  + \left[1-\frac{1}{2}{\rm erfc}\left(\frac{z-\gamma_0\sigma^2}{\sqrt{2}\sigma}\right)\right]
                    \exp\left(-\gamma_0z+\frac{\gamma_0^2\sigma^2}{2}\right),
\end{eqnarray*}
with $z=\rho-\rhosb$. In the last line, the first term comes from the event-by-event fronts which are at saturation ($T=1$) while the second one comes from the exponential tail. Practically, there are two interesting limits to analyse in more details:
\begin{enumerate}
\item \underline{$\gamma_0\sigma^2 \ll 1$}\\
When the dispersion is small (compared to $1/\gamma_0$ which is the typical decay length of the exponential tail), the dispersion can be neglected. We thus recover 
\begin{equation}\label{eq:avgt_gs}
\avg{T(\rho)} = T_{\text{event}}(\rho-\rhosb),
\end{equation}
and geometric scaling is satisfied. Since $\sigma^2\propto Y$, this is what happens in early stages of the evolution.
\item \underline{$\gamma_0\sigma^2 \gg 1$}\\
For large dispersion, {\em i.e.} at very high energies, we find
\begin{equation}\label{eq:avgt_ds}
\avg{T(\rho)} = \frac{1}{2}{\rm erfc}\left(\frac{z}{\sqrt{2}\sigma}\right).
\end{equation}
This expression is valid for $-\gamma_0\sigma^2 \ll z \ll \gamma_0\sigma^2$ which is arbitrarily large at high energy. 
\end{enumerate}

From this analysis, we deduce two very important physical consequences for the high-energy behaviour of the amplitudes:
\begin{itemize}
\item a new scaling law, called {\em diffusive scaling}, appears. The amplitude scales like $z/\sigma$, which, in more physical variables means that it only depends on the ratio $\log(r^2Q_s^2)/\sqrt{Y}$. It is possible to show that this property of diffusive scaling extends to physical observables such that DIS, diffractive DIS \cite{himst,cyr} and gluon production \cite{ims,cyr}.
\item The error function in \eqref{eq:avgt_ds} fully comes from the event-by-event fronts which are at saturation. In other words, the high-energy behaviour of the amplitudes can be obtained by taking $T_{\text{event}} = \theta(\rho_s-\rho)$. Thus, at high energy, the scattering amplitudes are dominated by {\em black spots}, $T=0$ or $T=1$.
\end{itemize}

\section{Discussion and perspectives}

Let us now summarise the main points raised in these proceedings. As this document is in itself some kind of a summary, I shall only pick up the most important points, refer to various references for detailed approaches and discuss open questions.

First, we have shown that it is possible to describe the evolution to high energy in pQCD by an infinite hierarchy of equations (see \eqref{eq:fluct}). This gives the evolution of the $\avg{T}$ matrix (averaged over the target wave-function) and its higher-order correlations $\avg{T^k}$. In this hierarchy, the evolution of $\avg{T^k}$ contains three types of contribution:
\begin{enumerate}
\item the linear BFKL growth, proportional to $\abar\avg{T^k}$. This is the usual high-energy contribution computed thirty years ago and corresponding to the exchange of $k$ pQCD Pomerons. It leads to a fast (exponential) increase of the scattering amplitude.
\item the saturation corrections, proportional to $\abar\avg{T^{k+1}}$. This negative term becomes important when the amplitude reaches unitarity and it allows for the constraint $T(\vvr,\vb)\le 1$ to be satisfied.
\item the fluctuations term, proportional to $\abar\alpha_s^2\avg{T^{k-1}}$. These fluctuations correspond to gluon-number fluctuations in the target. They play an important role when the amplitude is of order $\alpha_s^2$ or, equivalently, when the dipole density is of order one, {\em i.e.} in the dilute regime where one indeed expects fluctuation effects to appear.
\end{enumerate}

If one neglects the fluctuation contributions, we recover the Balitsky hierarchy (in its large-$N_c$ formulation) and, in the mean field approximation, the BK equation is obtained. This last equation, much simpler than the infinite hierarchy captures many important points concerning the physics of saturation. We also have to point out that, as long as we do not take into account fluctuations, the evolution at all orders in $1/N_c$ exists under the form of the Balitsky/JIMWLK equation. The Balitsky equations take the form of a hierarchy, similar to the one introduced in these proceedings but involving, in addition to dipoles, more complicated objects like quadrupoles, sextupoles, ... The JIMWLK equation is an equivalent formulation based on the Colour Glass Condensate approach, in which the probability density for the colour field of the target evolves through a functional equation. Further information concerning the Balitsky (resp. CGC) approach can be found in \cite{wilson} (resp. \cite{cgc}).

Many additional things can be said concerning those different contributions, {\em e.g.} concerning the equivalence between descriptions from the target and projectile point of view (see {\em e.g.} \cite{ddd}), or the attempts to describe the evolution beyond its large$-N_c$ limit \cite{beyond}. Those questions are active fields of research and the interested reader is forwarded to the corresponding references for further information.

In the second part of this overview of high-energy QCD, we have concentrated ourselves on the physical consequences arising from saturation and fluctuation effects. We have investigated how those contributions manifest themselves on the scattering amplitude and change the exponential behaviour obtained from the BFKL evolution. Again, the most important points can be summarised in two important steps which have their respective physical consequences:
\begin{enumerate}
\item Considering BFKL with saturation effects (mean-field picture), one can show that the BK equation lies in the same universality class as the FKPP equation. This implies that travelling waves are formed during the evolution towards high energy. From the BFKL kernel one obtains the critical parameters which are the anomalous dimension observed in the large-$Q^2$ tail and the speed of the wave, equivalent to the exponent of the saturation scale. In physical terms, these travelling waves correspond to geometric scaling. The experimental observation of geometric scaling in DIS at HERA can thus be seen as a consequence of saturation. It is extremely important to realise that geometric scaling is a prediction from saturation physics which extends to large values of $Q^2$, far beyond the saturation scale itself\footnote{The geometric scaling window grows, in logarithmic units, like $\sqrt{Y}$ beyond the saturation scale.}. This geometric scaling has also been predicted at nonzero momentum transfer, a result which may be applied for example to vector-meson production or DVCS.
\item If one includes the effects of fluctuations, the evolution becomes a Langevin equation equivalent to the stochastic FKPP equation, including a noise term. For each realisation of this noise, we observe geometric scaling with a decrease of the speed w.r.t. the mean-field case. If we consider a bunch of events, we observe that, although they all have the same shape, their position is diffused with diffusion increasing like $\sqrt{Y}$. This dispersion of the saturation scale for a set of events implies that geometric scaling is violated for the averaged amplitude. Recent studies indicate that geometric scaling is still valid at small rapidities. At higher energies, a new (diffusive) scaling is predicted \cite{himst} in which the amplitude scales w.r.t. the scaling variable $\tau=\log[k^2/Q_s^2(Y)]/\sqrt{Y}$.
\end{enumerate}

Although these two pictures endow most of the physical effects, they are far from being completely understood. Among the improvements, one can quote phenomenological applications like geometric and diffusive scaling predictions for vector-meson productions, for diffraction \cite{himst} and for LHC physics \cite{ims}. On more theoretical grounds most of the work that still need to be done concerns the effects of fluctuations. Most of the results known so far are derived within the local approximation for the noise term and neglecting the impact parameter dependence. In addition, apart from numerical studies, analytical results can only be applied to irrealistically small values of $\alpha_s$ and a better understanding for more physical values is still lacking. Although the existing picture is believed to contain all the qualitative effects, a more precise treatment is certainly an interesting and challenging problem to address. 

Within the framework presented here, the link between the QCD evolution equations and equations from statistical physics has led to a large number of interesting ideas and results. Even in statistical physics, many questions are yet opened, hence, extending the QCD picture including more sophisticated treatments of the BFKL kernel and more precise forms of the noise term is certainly not a straightforward task. The questions of making clear predictions for phenomenology, of understanding the effects of fluctuations beyond the local-noise approximation and of including impact-parameter dependence are hence expected to give interesting work in the near future. Among particle physics, high-energy QCD is thus one of the most active fields. Because of those open questions and because of the requirement for an excellent knowledge of QCD at the LHC, we expect that high-energy QCD is going to be even more active in the near future.

\begin{center}
{\bf Acknowledgements}
\end{center}
I would like to thank warmly Michal Praszalowicz and Andrzej Bialas, for the invitation to give a course in the Cracow School of Theoretical Physics. I also thank Robi Peschanski for a careful reading of this proceedings. G.S. is funded by the National Funds for Scientific Research (FNRS), Belgium.


\appendix

\section{BFKL evolution of dipole wavefunction}\label{app:bfkln}

Let us show that it is also possible to obtain the BFKL equation by looking at the evolution of the dipole densities in the target. We hence need to compute how, within a small increase of rapidity, the density of dipoles of coordinates $(\vx,\vy)$ evolves. The answer is
\[
\partial_Y n_{\vx\vy} = \int d^2z\,{\cal{M}}_{\vx\vz\vy} n_{\vx,\vz} + {\cal{M}}_{\vz\vy\vx} n_{\vz,\vy} - {\cal{M}}_{\vx\vz\vy} n_{\vx,\vz}
\]
The three contribution in this equation has well determined origin: the first term correspond to the situation where, at rapidity $Y$, we had a dipole $(\vx,\vz)$ which splits into $(\vx,\vy)$ and $(\vy,\vz)$ hence the positive contribution to the evolution of $n_{\vx\vy}$. The second term has similar explanation with an original dipole $(\vz,\vy)$ and the third one corresponds to the situation where we had one dipole $(\vx,\vy)$ at rapidity $Y$ which disappears through splitting into $(\vx,\vz)$ and $(\vz,\vy)$, leading to a loss term coming with a minus sign.

Getting to the the evolution of the scattering amplitude when we probe that system with a dipole $(\vu,\vv)$ required a few technical manipulations. First, we shall use eq. \eqref{eq:ntoT} to construct the scattering amplitude from the dipole density by convolution with the fundamental dipole-dipole interaction\footnote{The precise form of $\cal{A_0}$ (see \eqref{eq:A0}) is not really important, we shall only ask it to be conformal invariant.}:
\[
\partial_Y\avg{T_{\vu\vv}} = \int_{\vx\vy\vz} {\cal{A}}_0(\vu\vv|\vx\vy)\left[{\cal{M}}_{\vx\vz\vy} n_{\vx,\vz} + {\cal{M}}_{\vz\vy\vx} n_{\vz,\vy} - {\cal{M}}_{\vx\vz\vy} n_{\vx,\vz}\right].
\]
The trick to recover \eqref{eq:bfkl} is to use conformal properties. For example, for the first term in the equation, we perform the following change of variables\footnote{We use the complex notations in which a two-dimensional vector $\vx$ is equivalent to the complex number $x_1+ix_2$.} for the integration over $\vy$:
\[
y \to y'=f(y) = \frac{ay+b}{cy+d}\qquad\text{with }
\begin{cases}
a=-d=xu-zv,\;\;\;c=u-v+x-z,\\
b=uvz-uvx+xzv-xzu,
\end{cases}
\]
which has the effect to interchange $x$ with  $u$ and $z$ with $v$. It thus implies
\begin{eqnarray*}
\int_\vy {\cal{A}}_0(\vu\vv|\vx\vy){\cal{M}}_{\vx\vz\vy} 
 & \to & \int_{\vy'} {\cal{A}}_0(f(\vu)f(\vv)|f(\vx)f(\vy)){\cal{M}}_{f(\vx)f(\vz)f(\vy)}\\
 &  =  & \int_{\vy'} {\cal{A}}_0(\vx\vz|\vu\vy'){\cal{M}}_{\vu\vv\vy'}.
\end{eqnarray*}
Carrying on with the $\vx$ and $\vz$ integration will give
\[
\int_{\vy'} {\cal{M}}_{\vu\vv\vy'}\avg{T_{\vu\vy'}}.
\]
Renaming $\vy'$ into $\vz$ and performing a similar treatment for the remaining terms, one recovers the BFKL equation \eqref{eq:bfkl}.

\section{Langevin equation and infinite hierarchy}\label{app:langevin}

In this appendix, we show that a Langevin equation with a noise term proportional to $\sqrt{u(x,t)}\eta(x,t)$ (with $\eta$ a Gaussian white noise), turns into the fluctuation term in the hierarchy for averaged amplitudes. To stay as close as possible to the problem we are concerned with, let us consider a sFKPP-like equation with one spatial dimension and a generic linear kernel:
\[
\partial_t u(x,t) = \chi(\partial_x) u(x,t) - u^2(x,t) + \sqrt{2\sigma\,u(x,t)}\,\eta(x,t),
\]
with $\avg{\eta(x,t)} = 0$ and $\avg{\eta(x_1,t_1)\eta(x_2,t_2)} = \delta(t_1-t_2)\delta(x_1-x_2)$.

As usual, the Langevin equation has to be taken with the Ito prescription. This means that we consider discrete time steps $\Delta$ and write
\[
\frac{u_{i+1}(x)-u_i(x)}{\Delta} = \chi(\partial_x) u_i(x) - u_i^2(x) + \sqrt{2\sigma\,u_i(x)}\,\eta_i(x),
\]
with the r.h.s. taken at time $i$ and the following noise correlators:
\[
\avg{\eta_i(x)} = 0\qquad\text{ and } \qquad \avg{\eta_{i_1}(x_1)\eta_{i_2}(x_2)} = \frac{1}{\Delta}\delta_{i_1i_2}\delta(x_1-x_2).
\]

We can then consider the time evolution of any functional $F[u_i(x)]$ of the event-by-event amplitude $u_i(x)$. By series expansion and using the Langevin equation, we have
\begin{eqnarray*}
F[u_{i+1}(x)] 
 & = & F[u_i(x)]\\
 & + &\int dx\, \frac{\delta F[u_i(x)]}{\delta u_i(x)} \, 
                             \Delta [\chi(\partial_x) u_i(x) - u_i^2(x) + \sqrt{2\sigma\,u_i(x)}\,\eta_i(x)]\\
 & + & \frac{1}{2} \int dx_1 dx_2\, \frac{\delta^2 F[u_i(x)]}{\delta u_i(x_1)\,\delta u_i(x_2)} \\
 && \phantom{1intdx_1dx_2\,} \Delta [\chi(\partial_{x_1}) u_i(x_1) - u_i^2(x_1) + \sqrt{2\sigma\,u_i(x_1)}\,\eta_i(x_1)]\\
 && \phantom{1intdx_1dx_2\,} \Delta [\chi(\partial_{x_2}) u_i(x_2) - u_i^2(x_2) + \sqrt{2\sigma\,u_i(x_2)}\,\eta_i(x_2)].
\end{eqnarray*}
We want to obtain the evolution of the average amplitude $\avg{F[u(x,t)]}$. For that, we take the average in the previous equation and use the correlators of the noise term:
\begin{eqnarray*}
\lefteqn{\frac{\avg{F[u_{i+1}(x)]}-\avg{F[u_i(x)]}}{\Delta}}\\
 & = & \int dx\, \avg{\frac{\delta F[u_i(x)]}{\delta u_i(x)} \, [\chi(\partial_x) u_i(x) - u_i^2(x)]}\\
 & + & \frac{1}{2} \int dx_1 dx_2\, \avg{\frac{\delta^2 F[u(x)]}{\delta u(x_1)\,\delta u(x_2)} \, \Delta 
                             \,2\sigma\,\sqrt{u_i(x_1)u_i(x_2)}\,\frac{1}{\Delta}\delta(x_1-x_2)},
\end{eqnarray*}
where we have only kept the leading terms in $\Delta$.

One can finally take the continuum limit for the time variable and obtain 
\begin{eqnarray*}
\lefteqn{\partial_t \avg{F[u(x,t)]}}\\
 & = &\int dx\, \avg{\frac{\delta F[u(x,t)]}{\delta u(x,t)} \, [\chi(\partial_x) u(x,t) - u^2(x,t)]
                 + \frac{\delta^2 F[u(x,t)]}{[\delta u(x,t)]^2} \, \sigma\,u(x,t)}.
\end{eqnarray*}

In particular, one can consider the case $F[u(x)] = u^k(x)$, from which one recovers the infinite hierarchy of equations
\begin{eqnarray*}
\lefteqn{\partial_t \avg{u^k(x,t)}}\\
 & = & k\,\avg{u^{k-1}(x,t)\chi(\partial_x)u(x,t)} 
     - k\,\avg{u^{k+1}(x,t)}
     + k(k-1)\sigma\,\avg{u^{k-1}(x,t)}.
\end{eqnarray*}
In this expression, we clearly see that the last term, responsible for fluctuations, directly comes from the noise term in the Langevin equation. The two remaining ones correspond to the Balitsky hierarchy {\em i.e.} linear evolution nd saturation corrections.

\section{Saturation scale at nonzero momentum transfer}\label{app:nonzero}

Let us show in more details how the reference scale for the saturation scale at nonzero momentum transfer moves from $t=q^2$ at large $t$ to a soft scale $k_0^2$ when $t$ becomes small. From the condition for travelling-waves formation, we need to satisfy the requirement that the amplitude in the dilute tail (described by the linear equation) can be expressed as a superposition of waves. Hence, our starting point is the generic solution for projectile-target scattering in the BFKL limit:
\[
T_{\text{lin}}(\vk,\vq) = \int \frac{d\gamma}{2i\pi} e^{\abar\chi(\gamma)Y}\, f^\gamma(\vk,\vq)\,\phi_0^\gamma(Q_T,\vq),
\]
where $\vk$ is the momentum of the projectile dipole, $\vq$ is the momentum transfer and $Q_T$ is a typical scale for the target. The crucial point at this stage of the analysis is that, in this momentum space representation, we have a factorisation between a target and a projectile contribution. Since we are interested in the dilute tail of the scattering amplitude, the scale of the projectile is bigger than $Q_s^2$, $t$ and $Q_T^2$. In that limit, we know the asymptotic behaviour of the BFKL eigenfunctions
\[
f^\gamma(\vk,\vq) \propto \left(\frac{k^2}{q^2}\right)^{-\gamma.}
\]
On the target side, two limits are possible depending on the ordering between $t$ and $Q_T^2$. Again, we can expand the BFKL eigenfunctions to get\footnote{The power $1-\gamma$ appearing when $Q_T\gg q$ instead of $\gamma$ when $k\gg q$ comes from conformal properties of the BFKL equation.} 
\begin{eqnarray*}
\phi_0^\gamma(Q_T, \vq) & \stackrel{Q_T\ll q}{\propto} & 1\\
                 & \stackrel{Q_T\gg q}{\propto} & \left(\frac{Q_T^2}{q^2}\right)^{\gamma-1}.
\end{eqnarray*}
Reinserting everything back into the expression for the amplitude, we obtain the required exponential behaviour in the two limits:
\[
T_{\text{lin}}(\vk,\vq) \propto \int \frac{d\gamma}{2i\pi} e^{\abar\chi(\gamma)Y-\gamma L} \qquad\text{with }
L=\begin{cases}
\log(k^2/q^2) & \text{if }q\gg Q_T\\
\log(k^2/Q_T^2) & \text{if }q\ll Q_T
\end{cases}.
\]
This means that the reference scale for the saturation scale is fixed by the hardest between the target soft scale $Q_T^2$ and the momentum transfer $t$ as anticipated from the beginning.

\end{document}